\documentstyle[aps,prb,floats,fixtab,multicol,epsf]{revtex}
\newlength{\colw}
\newlength{\lshift}
\begin{document}
\draft
\title{Water Chemisorption and Reconstruction of the MgO Surface}
\author{K. Refson, R. A. Wogelius and D. G. Fraser}
\address{Department of Earth Sciences, Parks Road, Oxford OX1 3PR}
\author{M. C. Payne and M. H. Lee and V. Milman\protect\thanks{Present
address: Molecular Simulations, 240/250 The Quorum, Barnwell Road,
Cambridge CB5 8RE, UK}}
\address{Cavendish Laboratory, University of Cambridge, Madingley
Road, Cambridge CB3 0HE}
\date{December 6, 1994}
\maketitle

\begin{abstract}

The observed reactivity of MgO with water is in apparent conflict with
theoretical calculations which show that molecular dissociation does
not occur on a perfect (001) surface.  We have performed {\em
ab-initio}\/ total energy calculations which show that a chemisorption
reaction involving a reconstruction to form a (111) hydroxyl surface
is strongly preferred with $\Delta E = -90.2$~kJ mol$^{-1}$.  We
conclude that protonation stabilizes the otherwise unstable (111)
surface and that this, not the bare (001), is the most stable surface
of MgO under ambient conditions.

\end{abstract}


\begin{figure*}[t]
\epsffile{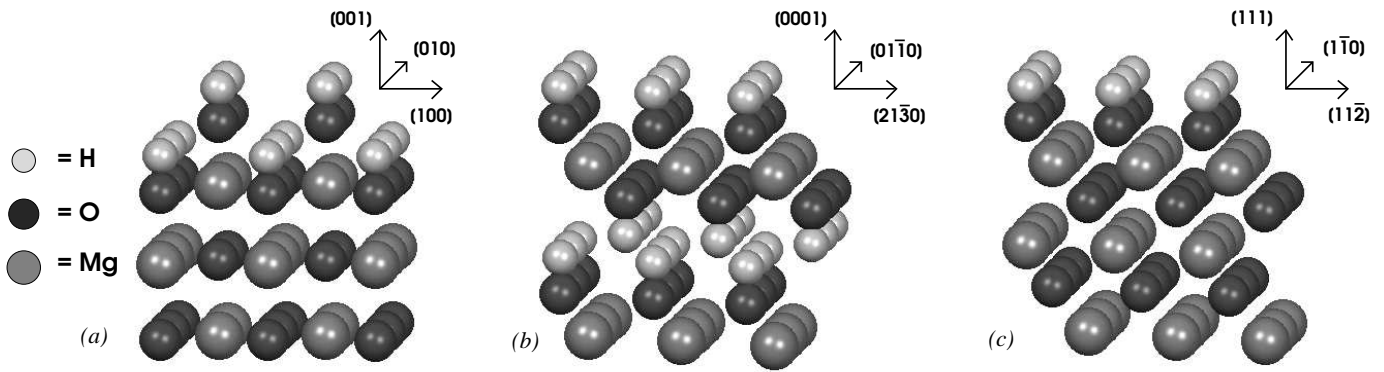}
\vspace{2mm}
\caption{(a) Hypothetical hydroxylated (001) surface of MgO, (b)
(0001) surface of Mg(OH)$_2$, (c) our postulated (111) hydroxyl
surface of MgO. The top three O-Mg-O layers of an oxygen-terminated
MgO (111) surface have the same structure as Mg(OH)$_2$
(0001). The (111) hydroxyl surface may be equivalently
constructed by protonation of an oxygen-terminated surface or by
hydroxylation of a magnesium-terminated surface.  }
\label{fig:mgosurf}
\end{figure*}

\begin{multicols}{2}

Magnesium oxide has long provided a prototype for the study of
surface structure and chemical reactions of oxides. Naturally occurring
MgO, known by its mineral name of periclase, is not a common
crustal mineral, but its simple structure makes it an excellent
example for the investigation of mineral surface chemistry.

Reactions at mineral surfaces are responsible for much of the chemical
change which occurs in the Earth's crust.  Weathering reactions
control the erosion of rocks and the consequent evolution of surface
topography thus providing an opposing mechanism to the more dramatic
process of mountain building.  Aqueous reactions in sedimentary basins
are responsible for the diagenetic processes which transform
unconsolidated sediments into rocks.  In this work we study the nature
of a simple mineral surface when exposed to an aqueous environment and
the chemical interaction of water with that surface. This is both a
prerequisite to studying the interaction with aqueous solutions and a
tractable first step towards ligand-exchange reactions in more complex
silicate minerals.

We have performed experiments on single-crystals of MgO prepared with
high-quality (001) faces which were reacted with acidic solutions. The
experiments and results are reported in detail
elsewhere\cite{wogelius:95}, the main feature being the development of
an altered surface layer. Elastic Recoil Detection Analysis
(ERDA)\cite{casey:88} shows protonation to a depth of 900 \AA\ with a
H/Mg ratio close to 2 giving a probable chemical composition of
magnesium hydroxide. Indeed brucite (the mineralogical name for
Mg(OH)$_2$) is the most common alteration product of periclase in the
natural environment\cite{dhz:66} and well-crystallized intergrowths of
brucite on periclase have been reported\cite{gorshkov:92}.

The initial stage in the reaction is hydroxylation of the surface. MgO
has the cubic rocksalt structure with (001) cleavage planes.  This is
the most stable surface and is the only one seen
experimentally\cite{colbourn:92}.  The simplest possibility for a
hydroxylated surface is obtained by dissociating a water molecule and
placing the OH group above each magnesium ion and the H above each
oxygen of the (001) surface (see Fig.~\ref{fig:mgosurf}a) as
postulated by Coluccia {\em et al.}\cite{coluccia:87}.

Some striking hydroxylation experiments were reported by Jones {\em et
al.\/} who studied surface roughening on (001) faces of
nanocrystalline MgO in a transmission electron
microscope\cite{jones:84}.  The remarkable affinity of MgO for water
is demonstrated by their {\em in situ}\/ observation of
hydration-induced surface roughening over 10 minutes under vacuum with
P$_{\text{H}_2\text{O}} < 10^{-5}$ Pa.  The presence of surface
hydroxyl groups on MgO powders exposed to H$_2$O has been demonstrated
by infra-red spectroscopy\cite{coluccia:87,kuroda:88,knozinger:93}.
Hydroxyls are clearly distinguishable from physisorbed molecular water
by the HOH bending mode which disappears above 100$^\circ$C, while the
OH stretching mode persists even above 500$^\circ$C.  Furthermore
there is complete monolayer coverage of the surface by hydroxyls, as
shown by microgravimetry measurements\cite{coluccia:87}.

Despite these observations, the most reliable theoretical calculations
predict that water molecules do not dissociate on the (001) surface.
Scamehorn {\em et} {\em al.}\cite{scamehorn:93} calculated the energetics of
the reaction \mbox{\{$>$MgO\} + H$_2$O $\rightleftharpoons$
\{$>$Mg(OH)$_2$\}}\cite{note:notation} to form the hydroxylated (001)
surface using periodic Hartree-Fock methods. They showed that
dissociative chemisorption is energetically unfavourable and that
physisorption of intact water molecules is preferred.  This was
confirmed by a Car-Parrinello {\em ab initio}\/ molecular-dynamics
study\cite{langel:94} which investigated the dynamics of a water
molecule at a MgO (001) surface. No dissociation occurred.
Experiments performed by Jones {\em et al.}\cite{jones:84} also led to
the conclusion that perfect (001) surface sites are not protonated.

In summary, water demonstrably chemisorbs onto MgO but trustworthy
calculations show that H$_2$O molecules should not dissociate on the
only known stable surface.   

Several authors have proposed that water
dissociates instead at low co-ordinated sites such as steps, corners
and other defects\cite{jones:84,onishi:87} and dissociation was
demonstrated computationally at steps and
corners\cite{langel:94,scamehorn:94}.  However one might expect this
to simply saturate the defect sites. Dissociation at defects does not
explain the monolayer coverage of the surface by OH
groups\cite{coluccia:87}, the whole-surface roughening observed in the
TEM\cite{jones:84} nor the progressive transformation of the entire
(001) surface and incipient bulk hydroxylation observed by i.r.\
spectroscopy\cite{jones:84,kuroda:88} and in our dissolution
experiments.

In order to account for these phenomena we propose an alternative
structure for the fully hydroxylated surface.
The progressive formation of a bulk hydrated layer suggests
consideration of the structural relationship between the oxide and the
crystalline hydroxide.  Mg(OH)$_2$ is trigonal and is composed of
layers of Mg$^{2+}$ and OH$^-$ ions in (0001) planes (see
Fig.~\ref{fig:mgosurf}b). This ``basal'' plane is also the cleavage
plane, yielding a stable type II hydroxide surface.  The trigonal
symmetry means that the basal plane is incommensurate with the cubic
(001) MgO surface, but closely resembles the threefold symmetric
(111).  Fig.~\ref{fig:mgosurf} illustrates that protonation of the
(111) creates a surface with the same structure as the Mg(OH)$_2$
(0001) cleavage plane.  This strongly suggests that protonation
may stabilize the MgO (111) surface, and that this may be the
hydroxylated surface observed by {\em i.r.}\/ spectroscopy.

Supporting evidence is provided by well-crystallized natural growths
of Mg(OH)$_2$ on MgO\cite{gorshkov:92} and from the inverse reaction,
the dehydration of Mg(OH)$_2$ to form MgO\cite{goodman:58}. In both
cases an epitaxial relationship exists between the two phases such
that the [0001] axis of the hydroxide is aligned with with the oxide
[111].

We have investigated the stability of the (111) hydroxyl surface using
total-energy pseudopotential calculations\cite{payne:92}.  The density
functional theory formulation of quantum mechanics is solved within
the local density approximation (using the parameterization of Perdew
and Zunger\cite{perdew:81}) by conjugate-gradient minimisation of the
total energy with respect to the valence electron wavefunctions.
Optimized non-local pseudopotentials are used in the Kleinman-Bylander
separable representation.  The system of ions and electrons is subject
to periodic boundary conditions which allows the use of a plane-wave
basis set.  The latter gives an accurate representation of the
crystalline environment. Just as importantly, it permits analytic
forces to be computed allowing for relaxation of the ions to their
minimum energy configuration.  This type of calculation has been used
to study defect energies in MgO\cite{devita:92}, OH groups as
substitutional defects in quartz\cite{lin:94}, reconstruction of the
silicon (111) surface\cite{stich:92} and dissociation
of Cl$_2$ at a silicon surface\cite{devita:93}.

One difficulty in using plane-wave pseudopotential methods for oxides
has been that the tightly-bound oxygen $2p$ electrons require a high
energy cutoff making the calculations expensive to perform.  Recent
advances have dramatically improved the convergence properties so that
oxide calculations are now routine\cite{lin:93}.  In this work we used
a new optimized oxygen pseudopotential which gives complete
convergence of the electronic energy (to 0.05 eV from an energy of
909 eV) at a cutoff of 500 eV.  Magnesium was also represented by an
optimized pseudopotential and hydrogen by a pure coulombic
potential.

As a check on the accuracy to be expected we performed calculations on
bulk MgO and an isolated water molecule, the initial reactants.  The
results agree closely with experimental values and are summarised in
table~\ref{table:bulk+h2o}.

\hspace{\lshift}
\begin{minipage}{\colw}
\begin{fixtable}
\begin{tabular}{lrrr}
 & Property & Calculated & Experimental \\ \hline
MgO & a$_0$(\AA) & 4.217$\pm$0.001 & 4.2117\cite{dhz:66} \\
    & K (Mbar)   & 1.62$\pm$0.02   & 1.603$\pm$0.003\cite{jackson:82} \\
    & K$^\prime$ & 4.19$\pm$0.08   & 4.15$\pm$0.10\cite{jackson:82} \\
H$_2$O & bond length (\AA) & 0.966 & 0.9572 \\
       & HOH angle ($^\circ$) & 103.9--104.1\tablenote{The
value depends on  the size and
shape of the supercell and the orientation of the
molecule, indicating a very small interaction between a
molecule and its periodic images.} & 104.52 \\
\end{tabular}

\caption{Calculated properties of bulk MgO and isolated water
molecule: MgO lattice parameter a$_0$, bulk modulus K and $K^\prime =
dK/dP$; H$_2$O molecular bond length and angle.}
\label{table:bulk+h2o}
\end{fixtable}
\end{minipage}

To represent a surface using periodic boundary conditions we performed
calculations on a slab of MgO in a periodic cell of
larger dimension leaving a vacuum separating periodic images of the
slab. The surface energy is given by \mbox{$\Delta E_{\text{surf}} =
(E_{\text{slab}} - E_{\text{bulk}})/a^2$.} To test convergence
with respect both to the number of layers in the slab and the vacuum
space we calculated the energy of the (001) MgO surface using 4, 6 or
8 atomic layers in the slab and different $c$ dimensions of the
supercell leaving between 8 and 16 \AA\ of vacuum.  The periodic cells
had dimensions $a_0/\sqrt{2} \times a_0/\sqrt{2} \times n a_0,
n=1,4,6$. The lattice parameter $a_0$ was fixed at the experimental
value of 4.2117 \AA\@.  The k point set contained 8 points for the
long cells and 32 or 48 points for the compact cell used for the bulk
calculation. In every case relative energies were computed using
equivalent sets to cancel basis-set size errors.  The total energies
were well-converged as a function of k-point sampling, to better than
0.05 eV out of 909 eV.  The co-ordinates of the ions in the outer
layers of the slab were relaxed to their minimum energy configuration,
while the inner 2 layers were constrained at their bulk separation.
The surface Mg$^{2+}$ ions moved inwards by 1\% and the
O$^{2-}$ ions by 0.1\% in agreement with previous Hartree-Fock
calculations\cite{causa:86}. The resultant surface energies are all
between 1.093 and 1.103 J m$^{-2}$ which shows that a 4-layer slab
with 8 \AA\ of vacuum gives energies converged to a precision of
better than 0.01 J m$^{-2}$. This result is compared with previous
work in table~\ref{table:surfens}.

\hspace{\lshift}
\begin{minipage}{\colw}
\begin{fixtable}
\begin{tabular}{lr}
Technique & $\Delta E_{\text{surf}}$ (J m$^{-2}$) \\ \hline
DFT/LDA this work & 1.10 \\
DFT/LDA using Gaussian basis set\cite{birkenheuer:94} & 1.16 \\
Periodic HF\tablenote{These calculations\cite{causa:86} used a basis
set containing only {\em s}\/ and {\em p}\/ orbitals. Birkenheuer
{\em et al.}\cite{birkenheuer:94} showed that including {\em d}\/ orbitals
decreased their $\Delta E_{\text{surf}}$ from 1.32 to
1.16 J m$^{-2}$.} & 1.43 \\
Pair Potentials\cite{mackrodt:88} & 1.07 \\
Experiment\tablenote{summarised by Tosi\protect\cite[and refs
therein]{tosi:64}.} & 1.04--1.20 
\end{tabular}

\caption{Calculated MgO(001) surface energies and comparison with
previous measurements and calculations.}
\label{table:surfens}
\end{fixtable}
\end{minipage}

The test calculations establish that errors in the energy
differences due to cell-size and basis-set effects are less
than 1~kJ mol$^{-1}$. This leaves one significant source of systematic
error, the LDA, which is known to underestimate molecular dissociation
energies\cite{devita:93}.  We estimated an upper bound for this error
of 40~kJ mol$^{-1}$ from a calculation of the reaction \mbox{MgO +
H$_2$O $\rightleftharpoons$ Mg(OH)$_2$}(brucite) whose energy is well
known . Our main result is quite robust to an error of this magnitude.

The chemisorption energy is fully determined given the structure and
state of the reactants and products by \mbox{$\Delta E =
E_{\text{product}} - \sum E_{\text{reactants}}$}.  The appropriate
initial states are the bare (001) surface of MgO and ice (since this
is a zero-temperature calculation) and the final structures are the
hydroxylated (001) and the (111) hydroxyl surfaces of
Fig.~\ref{fig:mgosurf}.  Scamehorn {\em et al.\/} showed that the
fully hydroxylated (001) surface has lower energy than if partially
hydrated\cite{scamehorn:93} so we performed calculations for only that
state.  Calculations on the hydroxylated (001) surface were performed
in a supercell $\sqrt{2} a_0 \times \sqrt{2} a_0 \times 4 a_0$
containing 4 layers of Mg and O ions plus 2 surface layers of hydrogen
or hydroxyl.  Simple electrostatic considerations indicate that full
hydroxylation must also be favoured for the (111) surface: this is the
only means of achieving a non-polar surface.

Ideally calculations on the reactants and products should use
equivalent supercells to achieve cancellation of basis set errors in
the computed $\Delta E$.  This rules out a direct calculation for the
bare (001) $\rightleftharpoons$ (111) hydroxyl reaction as the
symmetries differ and no cell can accommodate both structures. However
the reaction may be split into two stages \mbox{\{$>$MgO\}$^{(001)}$
$\rightleftharpoons$ MgO$^{\text{bulk}}$} and
\mbox{MgO$^{\text{bulk}}$ + H$_2$O $\rightleftharpoons$
\{$>$Mg(OH)$_2$\}$^{(111)}$} whose partial energies sum to the desired
result.  Each stage can be computed using cells of the appropriate
symmetry since bulk MgO is commensurate with both. The first stage is
simply the surface energy. The second used trigonal cells with $a =
a_0/\sqrt{2}, c=\sqrt{3} a_0$ and 24 k-points for bulk MgO and $a =
a_0/\sqrt{2}, c=2 \sqrt{3} a_0$ with 12 k-points and contained 5
layers of Mg and O ions in a hydroxylated slab. 
The outer layers were relaxed in all these calculations.

The initial state of water was based on calculations of an
isolated molecule in supercells equivalent to those used for the bulk
MgO calculations but of twice the linear dimensions to minimize
interaction between periodic images.  To this we added the experimental
value for the sublimation energy of ice,
$-47.35\pm0.02$~kJ mol$^{-1}$\cite{eisenberg:69}.

The results are listed in table~\ref{table:ohsurf}. As expected the
hydroxylation of the (001) surface is energetically unfavourable.  Our
hypothesis that hydroxylation stabilizes the (111) surface is
confirmed, with a $\Delta E$ of -90.2~kJ mol$^{-1}$ with respect to
ice, or (by using the results of Scamehorn {\em et
al.}\cite{scamehorn:93} for the physisorption energy) \mbox{-117.3~kJ}
mol$^{-1}$ relative to physisorbed water.  This provides a ready
explanation of the spectroscopic observations of surface hydroxyls and
the observed reactivity of MgO with water manifested as rapid
roughening of the (001) surfaces of microcrystals.  The chemisorption
reaction must involve a reconstruction of the (001) surface.

\hspace{\lshift}
\begin{minipage}{\colw}
\begin{fixtable}
\begin{tabular}{llrr}
Initial & Final & $\Delta E$ (J m$^{-2}$) & $\Delta E$ (kJ mol$^{-1}$) \\ \hline
(001) & (001) & $+0.78$ & $+41.6$\\
bulk  & (111) & $-0.59$ & $-31.3$ \\
(001) & (111) & $-1.69$ & $-90.2$\\
\end{tabular}
\caption{Computed energies for the chemisorption reaction of water
with MgO, formally \{$>$MgO\}$^{\text{initial}}$ + H$_2$O
$\protect\rightleftharpoons$ \{$>$Mg(OH)$_2$\}$^{\text{final}}$.  The
initial states are bulk MgO or the (001) surface plus ice at 0 K.  The
final states are the hydroxylated (001) or (111) hydroxyl surfaces.
The (111) surface energies are expressed per unit of the {\em
original\/} (001) surface area assuming a ratio of $\protect\sqrt{3}$:1.
A periodic HF study found a significantly more positive $\Delta
E$ for the hydroxylation of (001) of 77--90~kJ
mol$^{-1}$\protect\cite{scamehorn:93}.  The measured enthalpy of water
chemisorption on MgO at 613 Pa and 543 K is in the range $-113$ to $-189$
kJ mol$^{-1}$\protect\cite{beruto:93}.  Approximate thermodynamic
corrections to $\Delta E$ at 0 K for the (111) hydroxyl give $\Delta
H_{543} \protect\approx$ $-138$~kJ mol$^{-1}$, consistent with those
experiments.}
\label{table:ohsurf}
\end{fixtable}
\end{minipage}

We conclude that the (111) hydroxyl, rather than the bare (001) is the
normal surface of MgO under ambient environmental conditions.  A (001)
MgO surface will chemisorb water and reconstruct except under
ultra-high vacuum or high temperature.  Dehydroxylation is
experimentally observed only under UHV\cite{onishi:87}: indeed
hydration reactions occur under lesser vacuum in the TEM at a partial
pressure of water of $< 10^{-5}$ Pa.  Upon heating dehydroxylation
begins at 200$^\circ$C but is gradual with residual hydroxyls
persisting until over 700$^\circ$C\cite{coluccia:87,knozinger:93}.

This solves a number of experimental puzzles. 1) The \mbox{H$_2
\rightleftharpoons$ D$_2$} exchange reaction is catalysed by MgO at
temperatures as low as 78 K\cite{boudart:72}. Structural surface protons
provide the necessary exchange site\cite{kunz:77}. 2) It may also resolve
discrepancies between experimental and theoretical energies of
adsorption of CO onto MgO\cite{nygren:94}. The
appropriate surface for molecular adsorption is
the (111) hydroxyl surface, not the bare (001). 3)  Vermilyea
showed that the dissolution rates of MgO and Mg(OH)$_2$ are identical
over the pH range 2--5\cite{vermilyea:69}, an observation easily
explained by the almost identical surface structures.

Identification of the actual reconstruction pathway is beyond the
scope of this paper, but a plausible mechanism must account for both
the dissociation of the water molecule and the transport of
surface Mg$^{2+}$ and O$^{2-}$ ions.  Previous {\em ab initio}\/
calculations have shown that water molecules dissociate at surface
defects, particularly steps and corners\cite{langel:94,scamehorn:94},
corroborating the observation that the reconstruction is more rapid if
the surface is damaged\cite{jones:84}. The activation barrier for ion
transport may be estimated given that the inverse reconstruction,
the annealing of \{100\} facets on a (111) surface, is observed to
occur at 1000K\cite{onishi:87}. The equivalent thermal energy is 12~kJ
mol$^{-1}$, rather less than the chemical energies of the
hydroxylation reaction.  It is also possible that the
barrier for Mg$^{2+}$ transport away from the defect site is reduced
by hydration with either the product OH$^-$ ion or an intact H$_2$O
molecule. This would expose low co-ordinated oxygen sites able to
dissociate incoming water molecules and thereby continue the process.

Stabilization of a type III polar surface by protonation or
hydroxylation to form a non-polar type II surface is unlikely to be
unique to MgO.  We would also expect reconstructive chemisorption to
occur in other fcc binary metal oxides such as NiO\@.

The consequences of the stability of the hydroxyl surface over the
bulk are substantial but rather harder to predict.  It does provide a
driving force for the layer hydration observed by us and others which
apparently leads to a topotactic oxide-hydroxide transformation.  However a
much better characterization of the hydroxylated layer is needed and
further studies to establish its composition and structure are under
way. Whatever its nature it is demonstrably a vital precursor stage in
the dissolution of magnesium oxide and may also prove relevant to
understanding recrystallization and precipitation reactions.

{\bf Acknowledgment:}
This work was supported by NERC under grants GR3/8970 and GR3/8310.

%\bibliography{mgo,dft}

\begin{thebibliography}{10}
\bibitem{note:notation}
The braces are used to denote a surface
species and the $>$ symbol indicates the bond to the substrate.
\end{thebibliography}
%\bibliographystyle{prsty}

\end{multicols}

\end{document}